\documentclass[journal]{IEEEtran}
\IEEEoverridecommandlockouts

\usepackage{cite}
\usepackage{amsmath,amssymb,amsfonts}
\usepackage{newtxmath}
\usepackage{algorithmic}
\usepackage{graphicx}
\usepackage{booktabs}
\usepackage{textcomp}
\usepackage{xcolor}

\definecolor{myorange}{RGB}{255, 166, 106}
\definecolor{mygreen}{RGB}{34, 177, 76}
\usepackage{soul}

\usepackage{tikz}
\usetikzlibrary{positioning, backgrounds}
\usepackage[caption=false,font=footnotesize]{subfig}

\tikzset{
  annot/.style={font=\scriptsize, text=white, fill=black!75, inner sep=2pt, rounded corners=1pt, align=center},
  arrow/.style={->, white, line width=1.2pt},
  arrowbg/.style={->, black!50, line width=2pt},
}

\usepackage[hyphens]{url}
\usepackage{breqn}

\usepackage{caption}
\usepackage{subcaption}
\usepackage{lettrine}

\usepackage{glossaries}

\makeglossaries

\newacronym{mmwave}{mmWave}{millimeter-wave}
\newacronym{siso}{SISO}{single-input single-output}
\newacronym{miso}{MISO}{multiple-input single-output}
\newacronym{simo}{SIMO}{single-input multiple-output}
\newacronym{mimo}{MIMO}{multiple-input multiple-output}
\newacronym{tx}{Tx}{transmitter}
\newacronym{rx}{Rx}{receiver}
\newacronym{rf}{RF}{radio frequency}
\newacronym{mpc}{MPC}{multipath component}
\newacronym{ula}{ULA}{uniform linear array}
\newacronym{aoa}{AOA}{angle-of-arrival}
\newacronym{aod}{AOD}{angle-of-departure}
\newacronym{cir}{CIR}{channel impulse response}
\newacronym{pa}{PA}{power amplifier}
\newacronym{if}{IF}{intermediate frequency}
\newacronym{lo}{LO}{local oscillator}
\newacronym{clk}{CLK}{clock reference}
\newacronym{dac}{D/A}{digital-to-analog converter}
\newacronym{zc}{ZC}{Zadoff-Chu}
\newacronym{fpga}{FPGA}{Field-Programmable Gate Array}
\newacronym{los}{LOS}{Line-Of-Sight}
\newacronym{nlos}{NLOS}{Non Line-Of-Sight}
\newacronym{hpbw}{HPBW}{half-power beam width}
\newacronym{subthz}{sub-THz}{sub-terahertz}
\newacronym{vaa}{VAA}{virtual antenna array}
\newacronym{dss}{DSS}{directional scanning scheme}
\newacronym{snr}{SNR}{signal-to-noise-ratio}
\newacronym{uca}{UCA}{uniform circular array}
\newacronym{ura}{URA}{uniform rectangular array}
\newacronym{sage}{SAGE}{space-alternating generalized expectation-maximization}
\newacronym{vna}{VNA}{vector network analyzer}
\newacronym{lna}{LNA}{low-noise amplifier}
\newacronym{pdp}{PDP}{power-delay profile}
\newacronym{pmf}{PMF}{polymer microwave fiber}
\newacronym{ap}{AP}{access point}
\newacronym{ru}{RU}{radio unit}
\newacronym{hrpe}{HRPE}{high-resolution parameter estimation}
\newacronym{rms}{RMS}{root mean square}

\begin{document}
\bstctlcite{BSTcontrol}

\title{Vertical Sub-THz Channel Characterization: Sounder Implementation and Initial Measurements
\thanks{

Ali Al-Ameri, Juan Sanchez, Aleksei Fedorov, Buon Kiong Lau, Ove Edfors and Fredrik Tufvesson are with the Department of Electrical and Information Technology, Lund University, 22354 Lund,
Sweden (e-mail:$\{$ ali.al-ameri; juan.sanchez; aleksei.fedorov; buon\_kiong.lau; ove.edfors; fredrik.
tufvesson$\}$@eit.lth.se).
Xuesong Cai is with the State Key Laboratory of Photonics and Communications, School of Electronics, Peking University, Beijing 100871, P. R. China
(e-mail: xuesong.cai@pku.edu.cn).

}
}

\author{Ali~Al-Ameri,~\IEEEmembership{Student member,~IEEE},
Xuesong~Cai,~\IEEEmembership{Senior member,~IEEE}, 
Juan Sanchez,~\IEEEmembership{Student member,~IEEE}, Aleksei Fedorov,
        Buon Kiong Lau,~\IEEEmembership{Fellow,~IEEE}, Ove Edfors,~\IEEEmembership{Senior member,~IEEE} 
  and~Fredrik~Tufvesson,~\IEEEmembership{Fellow,~IEEE}}

\markboth{Vertical Sub-THz Channel Characterization:
Sounder Implementation and Initial Measurements}%
{Al-Ameri \MakeLowercase{\textit{et al.}}: Vertical Sub-THz Channel Characterization}

\maketitle 

\begin{abstract} 
We present a measurement-based characterization of indoor vertical ceiling-to-ground sub-THz channels in the 136--144~GHz band, motivated by ceiling-mounted radio-unit deployments for future distributed indoor networks. The measurements are performed using a vector network analyzer (VNA)-based channel sounder with a mechanically scanned planar virtual antenna array (VAA) at the receiver, enabling single-input single-output (SISO), small-array single-input multiple-output (SIMO), and large-array SIMO measurements in three representative indoor environments: an office, a laboratory, and a ventilation room. The small-array and large-array SIMO measurements synthesize $2\times2$~cm and $30\times1$~cm uniform rectangular arrays (URAs), respectively. The results show that the vertical links are generally dominated by a strong Line-of-Sight (LOS) component close to the ceiling to ground direction, but with clear differences between the different environments. The office and laboratory exhibit a relatively limited delay dispersion, whereas the ventilation room shows stronger delayed multipath components due to its corrugated metallic ceiling and surrounding metallic structures. The measured root mean square (RMS) delay spreads are $0.55$--$1.74$~ns for the small-array measurements and $0.44$--$2.57$~ns for the large-array measurements, which are smaller than those reported in several horizontal indoor sub-THz measurement campaigns at similar frequencies. However, the channel is not purely free-space. Repeatable second-order reflections present in all of the environments involving the receiver table, ceiling, transmitter structure, and ceiling-mounted objects are observed. The large-array measurements further reveal spatial non-stationarity along the 30~cm aperture, with several multipath components visible only over limited parts of the array. The results indicate that ceiling-mounted sub-THz links can benefit from short vertical propagation distances and limited delay dispersion, but ceiling materials, suspended obstructions, and aperture-dependent channel variations must be considered in channel modeling and deployment planning.

\end{abstract}
\begin{IEEEkeywords}
Channel sounder design, dynamic channels, sub-THz, propagation measurements, ceiling-mounted access points.
\end{IEEEkeywords} 

\IEEEpeerreviewmaketitle

\section{Introduction} \label{sec:introduction}

\lettrine[lines=2]{T}{he} 
 increasing demand for extremely high data rates and accurate localization in future wireless systems has motivated the exploration of \gls{mmwave} and \gls{subthz} frequency bands, where large frequency bandwidths are available \cite{8732419,10439212}. In particular, indoor sub-THz communication is considered a key enabler for data-intensive applications envisioned for sixth-generation (6G) networks, such as extended reality, industrial automation, and high-capacity wireless access \cite{9144301}. However, operation at these frequencies introduces significant propagation challenges, including high path loss, limited diffraction, and increased sensitivity to blockage \cite{9769234,9838910}.

\begin{figure}[h]

    \centering
    \includegraphics[width=0.8\columnwidth]{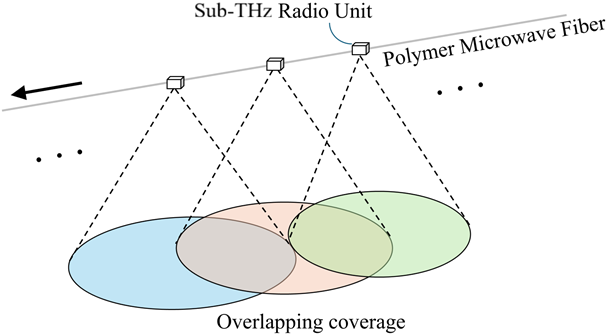}

    \caption{Ceiling-mounted distributed sub-THz architecture with PMF-interconnected RUs providing short-range vertical links.}
    \label{fig:6gtandem}
\end{figure}

One promising approach to mitigate these challenges in indoor environments is the distributed architecture proposed in the 6GTandem project \cite{11225925}. In this architecture, compact sub-THz \glspl{ru} are mounted on the ceiling and interconnected via \glspl{pmf} \cite{strombeck2023}, enabling short-range vertical connections to user devices below, as illustrated in Fig.~\ref{fig:6gtandem}. In such ceiling-mounted configurations, the propagation geometry differs fundamentally from conventional horizontal links, as the channel is dominated by vertical propagation and strong interactions with the ground plane and surrounding surfaces such as tables. In particular, reflections from the ground and ceiling may contribute significantly to the received signal power, with direct implications for array processing and equalizer design at sub-THz frequencies.

Despite the growing interest in indoor sub-THz systems, experimental characterization of vertical indoor channels remains limited. Existing sub-THz channel measurements predominantly focus on horizontal links \cite{9838910,9500596,11000040} and typically rely on \glspl{dss} \cite{xing2019indoorwirelesschannelproperties} or \glspl{vaa} with circular geometries \cite{10621601,10201930}. While these approaches provide valuable insight into angular characteristics, they are not well suited for capturing the spatial behavior of vertically oriented channels, where planar sampling and aperture extension along the ground plane are essential. As a result, key aspects such as the strength of ground-bounce components, the spatial structure of the channel, spatial non-stationarity along large-arrays, and the impact of ceiling-mounted obstructions are not yet well understood in the sub-THz band.

In this paper, we present a measurement-based characterization of the indoor vertical sub-THz channel in the 136--144~GHz band. The work is enabled by a \gls{vna}-based channel sounder with a planar \gls{vaa} positioned on a horizontal surface, allowing spatial sampling of the received field in a ceiling-to-ground geometry. Measurements are conducted in three representative indoor environments using \gls{siso} and \gls{simo} configurations with different virtual array apertures. The results are used to analyze large-scale power decay, spatial multipath structure, and non-stationarity along extended apertures, with emphasis on propagation mechanisms relevant to ceiling-mounted sub-THz deployments.

The paper is organized as follows. Section~II describes the channel sounder implementation and measurement setup. Section~III presents the measurement campaign and results, and Section~IV concludes the paper.

\section{Channel sounder implementation and measurement setup}
\label{sec:SounderImplementationnAndPractical}

\subsection{Channel Sounder Implementation}

Channel measurements were conducted in the D-band, covering the range from 136 to 144~GHz. To this end, a \gls{vna}-based channel sounder configured as a \gls{vaa} was employed.

\begin{figure}[h]
    \centering
    \includegraphics[width=01\linewidth]{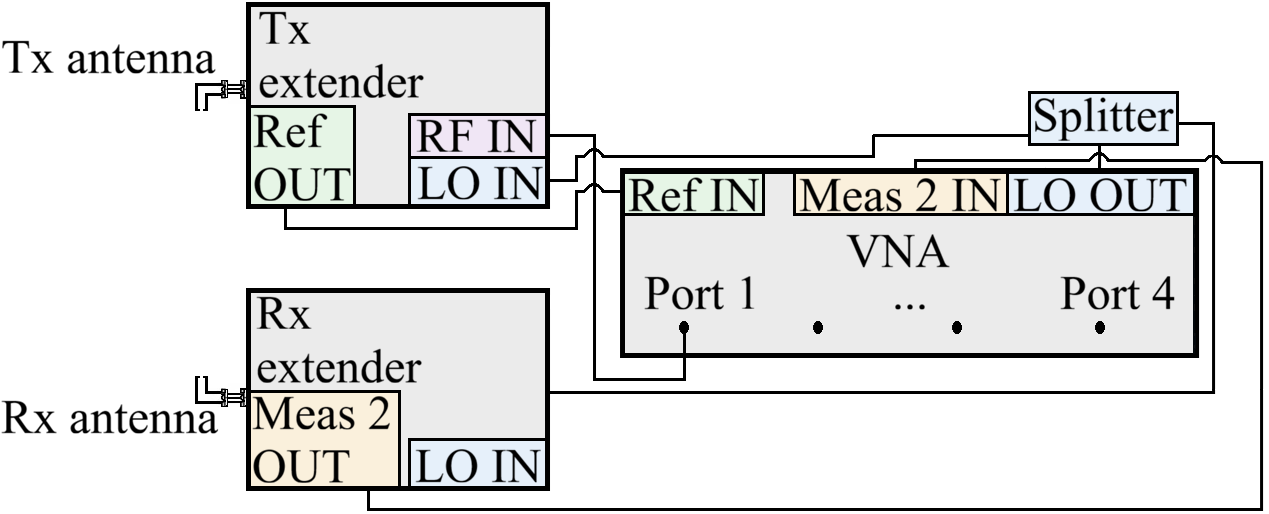}
    \caption{Channel sounder block diagram.}
    \label{fig:blockdiagram}
\end{figure}

The sounder is built around an R\&S ZNA43 \gls{vna}, which performs complex $S_{21}$ measurements up to 43~GHz. Frequency conversion to the target sub-THz band is achieved using R\&S ZC170 frequency extenders operating as \gls{tx} and \gls{rx} modules. A block diagram of the sounder architecture is shown in Fig.~\ref{fig:blockdiagram}. The system operates by measuring the frequency response between the \gls{tx} and \gls{rx} ports, with the extenders providing upconversion and downconversion through integrated multipliers, mixers, and filters.

Specifically, the \gls{if} signal from \gls{vna} Port~1 is routed to the transmitter-side extender, where it is upconverted to the desired \gls{rf} frequency using a common \gls{lo} shared between the \gls{tx} and \gls{rx}. A portion of the signal is coupled internally within the extender and returned to the \gls{vna} reference input to ensure phase coherence. After propagation through the wireless channel, the received signal is downconverted by the \gls{rx} side extender and fed back to the \gls{vna} measurement port, where the complex channel frequency response is obtained.

A limitation of this configuration is the restricted measurement distance imposed by losses in the interconnecting cables between the \gls{vna} and the frequency extenders. This effect is particularly pronounced due to the relatively high \gls{if} (11.33--12~GHz) and \gls{lo} (13.57--14.37~GHz) frequencies required for operation in the 136--144~GHz band, which result in non-negligible cable attenuation and reduced dynamic range. While techniques such as radio-over-fiber \cite{10621601,8901446,10999526} or the inclusion of sub-THz waveguide \glspl{lna} and \glspl{pa} could be employed to mitigate these limitations, they were not required in this work since the considered vertical ceiling-to-ground measurement distances are relatively short.

To enable spatial characterization of the channel, a \gls{vaa} was implemented at the \gls{rx} side for \gls{aoa} estimation. While conventional \gls{vaa} realizations typically involve moving only the antenna element, operation at sub-THz frequencies requires waveguide-fed antennas directly connected to the frequency extender. Consequently, the entire receiver, including the receive antenna and the receiver-side frequency extender, must be mechanically moved.

\begin{figure}[h]
  \centering
  \begin{tikzpicture}
    \node[anchor=south west, inner sep=0] (img) 
      {\includegraphics[width=\columnwidth]{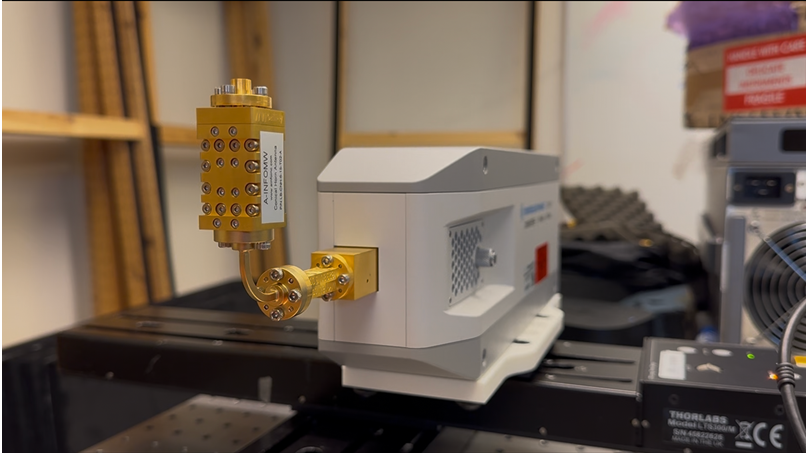}};
    \begin{scope}[x={(img.south east)}, y={(img.north west)}]

      \node[annot] (rxant) at (0.10, 0.75) {Rx antenna};
      \draw[arrow] (rxant.east) -- (0.25, 0.7);
      
      \node[annot] (rxfe) at (0.55, 0.8) {Rx frequency extender};
      \draw[arrow] (rxfe.south) -- (0.55, 0.67);
      
      \node[annot] (plat) at (0.85, 0.35) {3D Printed\\supporting platform};
      \draw[arrow] (plat.west) -- (0.65, 0.23);
      
      \node[annot] (stage) at (0.15, 0.20) {Motorized\\linear stages};
      \draw[arrow] (stage.east) -- (0.32, 0.18);
      \draw[arrow] (stage.east) -- (0.4, 0.05);

    \end{scope}
  \end{tikzpicture}
  \caption{Receiver setup with frequency extender mounted on motorized linear stages.}
  \label{fig:receiverarray}
\end{figure}
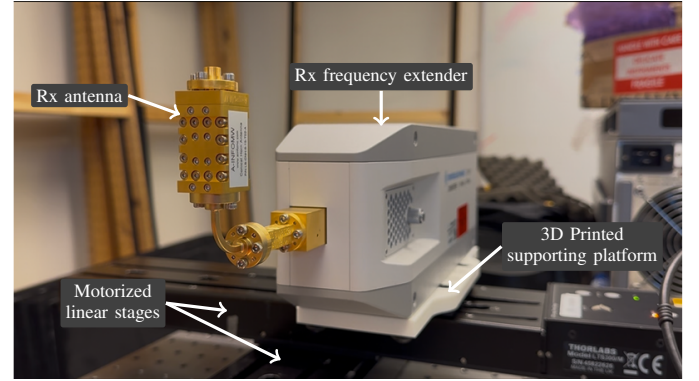

The \gls{rx} assembly was mounted on a 3D-printed support structure attached to two orthogonally stacked 300~mm motorized linear stages positioned on a table. This configuration enables controlled two-dimensional scanning of the receiver over a rectangular grid along the ground plane, thereby synthesizing a planar \gls{ura}, as illustrated in Fig.~\ref{fig:receiverarray}. Notice the receive antenna is oriented to point towards the ceiling via a 90$^\circ$ waveguide bend.

Phase stability is a major difficulty when working with these frequencies. At the highest frequency in the swept range of 136-144 GHz, the wavelength is around 2 mm. This means that if the positioners moving the \gls{rx} converter and the antenna connected to create the \gls{vaa}, move a fraction of a millimeter too much or too little, they would render the \gls{vaa} unreliable for directional estimation. To minimize this, we used Thorlabs LTS300C motorized linear stages, which have a bidirectional repeatability	$\le$±2 µm. Furthermore, the scanning grid and speed, as well as acceleration, were chosen to minimize movement inaccuracies.

\subsection{Measurement setup}

Measurements were conducted in three different indoor environments. In each environment, three measurement scenarios were considered, except for the office environment where only \gls{siso} and small-array \gls{simo} measurements were performed. Each scenario was designed to investigate a distinct aspect of the vertical ceiling-to-ground sub-THz channel.

First, \gls{siso} measurements were carried out to characterize the large-scale power decay and path loss behavior across the different environments. 
The measurements provide insight into the attenuation associated with repeated ceiling–ground reflections and the strength of higher-order bounce components.

Second, small-array \gls{simo} measurements using a $20 \times 20$ element virtual array were performed to analyze the spatial characteristics of the channel. These measurements enable the identification of dominant multipath components, their angular distribution at the receiver, and the relative contribution of the \gls{los} and reflected paths.

Finally, large-array \gls{simo} measurements employing a $300 \times 10$ element virtual array were conducted to investigate spatial non-stationarity along the array aperture. This configuration allows assessing variations in multipath structure across the array 
and the impact of partial blockage by ceiling-mounted obstructions on the received signal at sub-THz frequencies.

\begin{figure}[h]
  \centering
  \subfloat[Close look to the system.]{
  \begin{tikzpicture}
    \node[anchor=south west, inner sep=0] (img) 
      {\includegraphics[width=.46\columnwidth]{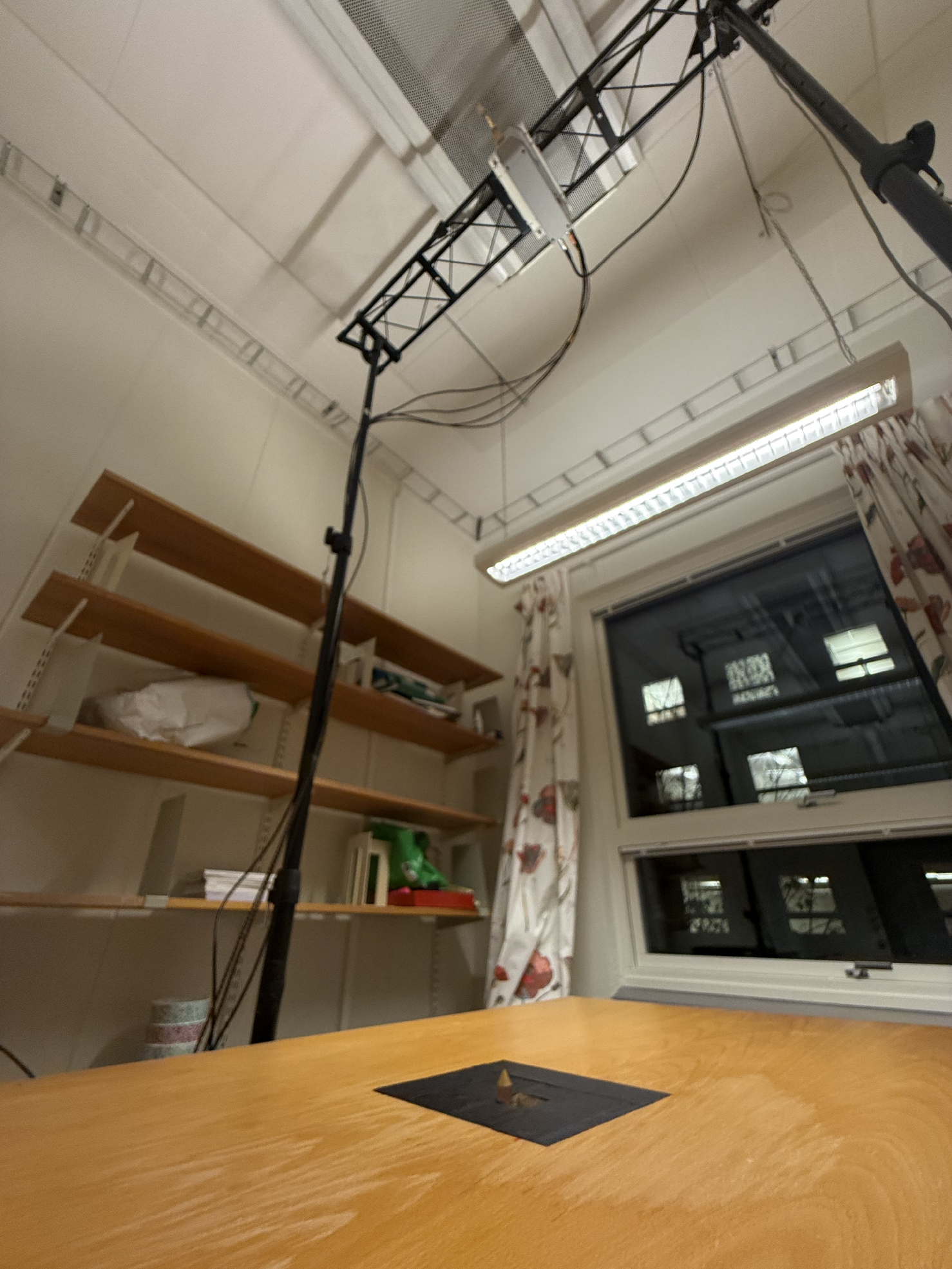}};
    \begin{scope}[x={(img.south east)}, y={(img.north west)}]
      \useasboundingbox (0,0) rectangle (1,1);

      \node[annot] (txant) at (0.25, 0.9) {Tx};
      \draw[arrow] (txant.east) -- (0.5, 0.91);
      
      \node[annot] (txfe) at (0.25, 0.75) {Tx frequency\\extender};
      \draw[arrow] (txfe.east) -- (0.56, 0.85);
      
      \node[annot] (pole) at (0.75, 0.45) {Metallic poles};
      \draw[arrow] (pole.north) -- (0.37, 0.6);
      \draw[arrow] (pole.north) -- (0.96, 0.82);
      
      \node[annot] (rxant) at (0.3, 0.1) {Rx};
      \draw[arrow] (rxant.east) -- (0.52, 0.14);

    \end{scope}
  \end{tikzpicture}
  \label{fig:measurement_setup}
  }
  \hfill
  \subfloat[Overall office environment.]
  {
  \begin{tikzpicture}
    \node[anchor=south west, inner sep=0] (img) 
      {\includegraphics[width=.46\columnwidth]{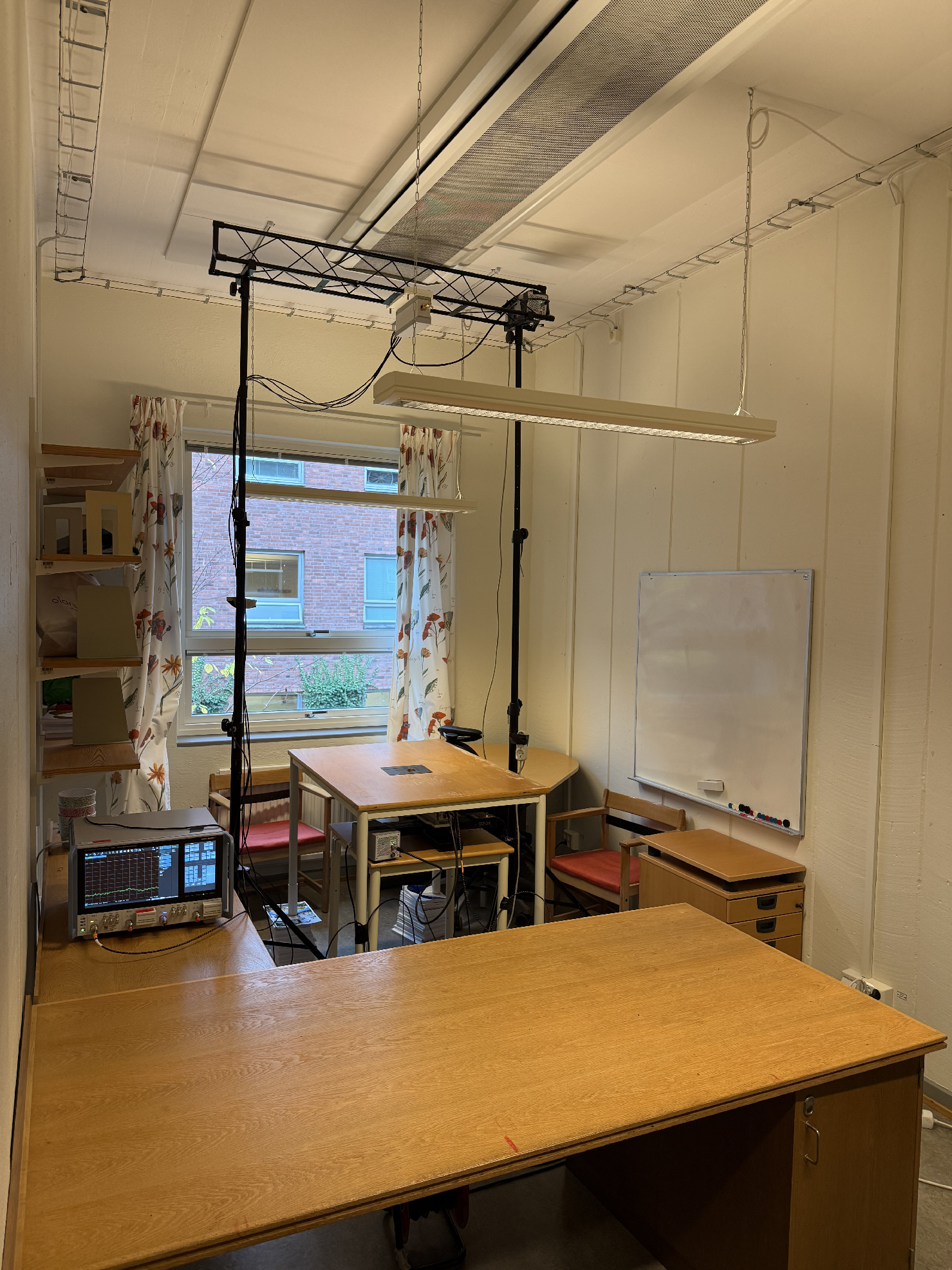}};
    \begin{scope}[x={(img.south east)}, y={(img.north west)}]
      \useasboundingbox (0,0) rectangle (1,1);

      \node[annot] (mesh) at (0.2, 0.9) {Metallic mesh};
      \draw[arrow] (mesh.east) -- (0.5, 0.9);
      
      \node[annot] (wallA) at (0.25, 0.7) {Wall A};
      \draw[arrow] (wallA.west) -- (0.05, 0.7);
      
      \node[annot] (shelves) at (0.25, 0.5) {Shelves};
      \draw[arrow] (shelves.north) -- (0.1, 0.64);
      \draw[arrow] (shelves.south) -- (0.1, 0.4);

      \node[annot] (vna) at (0.25, 0.15) {VNA};
      \draw[arrow] (vna.north) -- (0.12, 0.28);

      \node[annot] (board) at (0.8, 0.6) {Whiteboard};
      \draw[arrow] (board.south) -- (0.75, 0.5);

      \node[annot] (desk) at (0.78, 0.32) {Shielding desk};

      \draw[arrow] (desk.west) -- (0.5, 0.38);

      \node[annot] (lamp) at (0.8, 0.8) {Ceiling lamps};
      \draw[arrow] (lamp.south) -- (0.75, 0.67);
      \draw[arrow] (lamp.south) -- (0.45, 0.61);

      \node[annot] (tx) at (0.5, 0.8) {Tx};
      \node[annot] (rx) at (0.48, 0.43) {Rx};

    \end{scope}
  \end{tikzpicture}
  \label{fig:enviornment_office}
  }
  \caption{Office measurement environment.}
  \label{fig:setup_combined}
\end{figure}
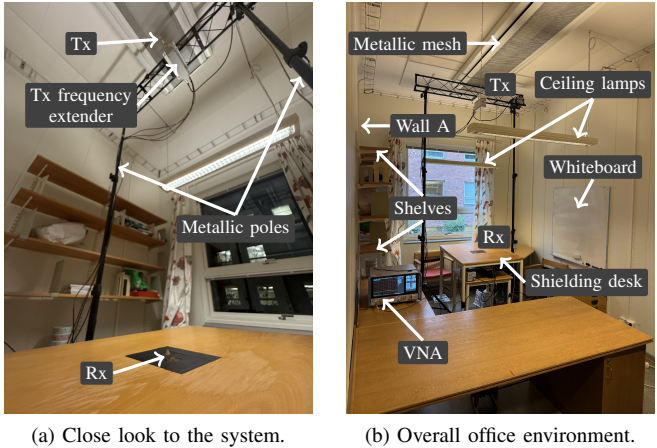

All measurements were conducted using the \gls{vna}-based \gls{vaa} sounder shown in Figs.~\ref{fig:blockdiagram} and~\ref{fig:receiverarray}. The setup consists of \gls{tx} and \gls{rx} frequency extenders connected to their respective antennas through waveguide interfaces. The transmitter-side extender was mounted on a rigid structure supported by metallic poles, positioning the \gls{tx} antenna near the ceiling and pointing downward toward the receiver. This configuration emulates a ceiling-mounted access point deployment.

For the \gls{siso} and small-array \gls{simo} measurements, a table surface was placed around the receiver so that only a $3\times3$~cm opening was exposed for the \gls{rx} antenna. The surrounding surface was shielded to reduce unwanted reflections from the mechanical positioners and supporting structures, as shown in Fig.~\ref{fig:measurement_setup}. The main configuration parameters for the different measurement scenarios are summarized in Table~\ref{table:parameters}.

\begin{table*}
\centering
\caption{Channel sounder configuration parameters for the SISO, small-array SIMO, and large-array SIMO measurements.}
\label{table:parameters}
\renewcommand{\arraystretch}{1.1}
\begin{tabular}{|c c c c|}
\hline
\textbf{Parameter} & \textbf{SISO} & \textbf{Small-array} & \textbf{Large-array } \\
\hline\hline
Sounder type & VNA based & VNA based VAA & VNA based VAA \\
Frequency sweep & 134--146 GHz & 134--146 GHz & 134--146 GHz \\
Number of frequency samples & 2001 & 1001 & 1001 \\
\gls{tx} antenna (Gain) & Horn antenna (17 dB) & Open waveguide (6 dB) &  Horn antenna (17 dB)\\
\gls{rx} antenna (Gain) & Open waveguide probe (6 dB) & Open waveguide probe (6 dB) & Open waveguide probe (6 dB) \\
\gls{tx} 3dB beamwidth E-/H-Plane  & 25$^\circ$/29$^\circ$ & 62$^\circ$/94$^\circ$& 25$^\circ$/29$^\circ$  \\
\gls{rx} 3dB beamwidth E-/H-Plane  &  62$^\circ$/94$^\circ$ & 62$^\circ$/94$^\circ$ & 62$^\circ$/94$^\circ$ \\
\gls{rx} array  & -- & 20 X 20 elements (2 by 2 cm) \gls{ura} & 300 X 10 elements (30 by 1 cm) \gls{ura}  \\

\hline
\end{tabular}
\end{table*}

\section{Measurement Campaign and Results}
\label{sec:verification_measurement_results}

\subsection{Measurement Environments}

\subsubsection{Office}

The first measurement environment was an office at the Department of Electrical and Information Technology (EIT), Lund University, as shown in Fig.~\ref{fig:setup_combined}. In this environment, \gls{siso} and small-array \gls{simo} measurements were performed at the location shown in the figure. The office represents a typical workspace characterized by concrete walls, wooden furniture, low-hanging ceiling-mounted lighting fixtures, and a suspended metallic mesh ceiling.

\subsubsection{Laboratory Environment}

\begin{figure}[hbt!]
  \centering
  \subfloat[Large-array \gls{simo} measurements: ceiling lamp creates obstruction.]{
  \begin{tikzpicture}
    \node[anchor=south west, inner sep=0] (img) 
      {\includegraphics[width=.46\columnwidth, viewport=0 970 3225 5309, clip]{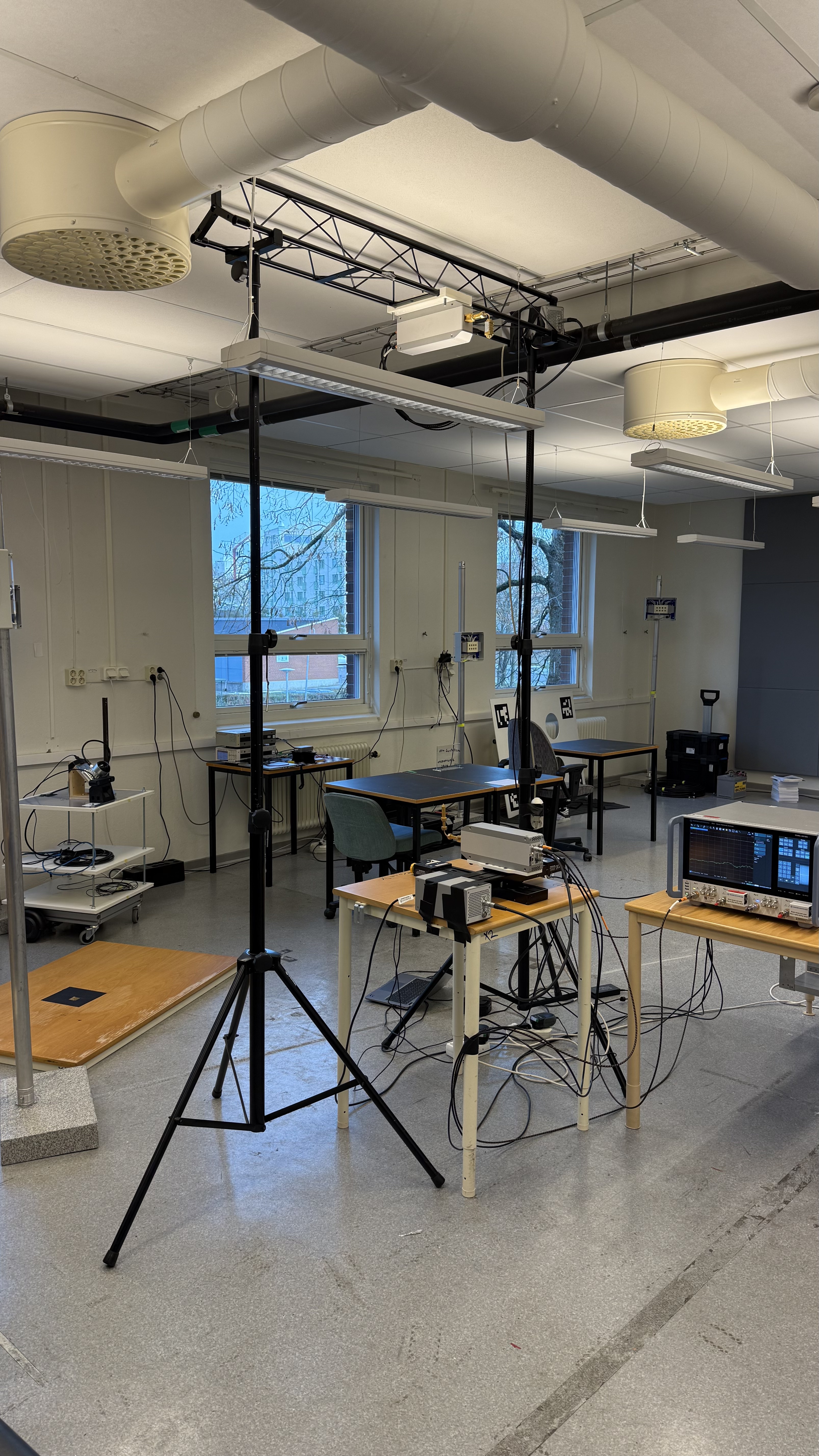}};
    \begin{scope}[x={(img.south east)}, y={(img.north west)}]
      \useasboundingbox (0,0) rectangle (1,1);

      \node[annot] (lamp) at (0.5, 0.6) {Ceiling lamp};
      \draw[arrow] (lamp.north) -- (0.5, 0.73);

      \node[annot] (vna) at (0.8, 0.2) {VNA};
      \draw[arrow] (vna.north) -- (0.9, 0.3);
      
      \node[annot] (tx) at (0.75, 0.85) {Tx};
      \draw[arrow] (tx.west) -- (0.605, 0.79);

      \node[annot] (rx) at (0.4, 0.4) {Rx};
      \draw[arrow] (rx.east) -- (0.53, 0.37);

    \end{scope}
  \end{tikzpicture}
  \label{fig:enviornment_5glab1}
  }
  \hfill
  \subfloat[Setup for \gls{siso} and small-array \gls{simo} measurements.]
  {
  \begin{tikzpicture}
    \node[anchor=south west, inner sep=0] (img) 
      {\includegraphics[width=.46\columnwidth, viewport=0 540 3300 5000, clip]{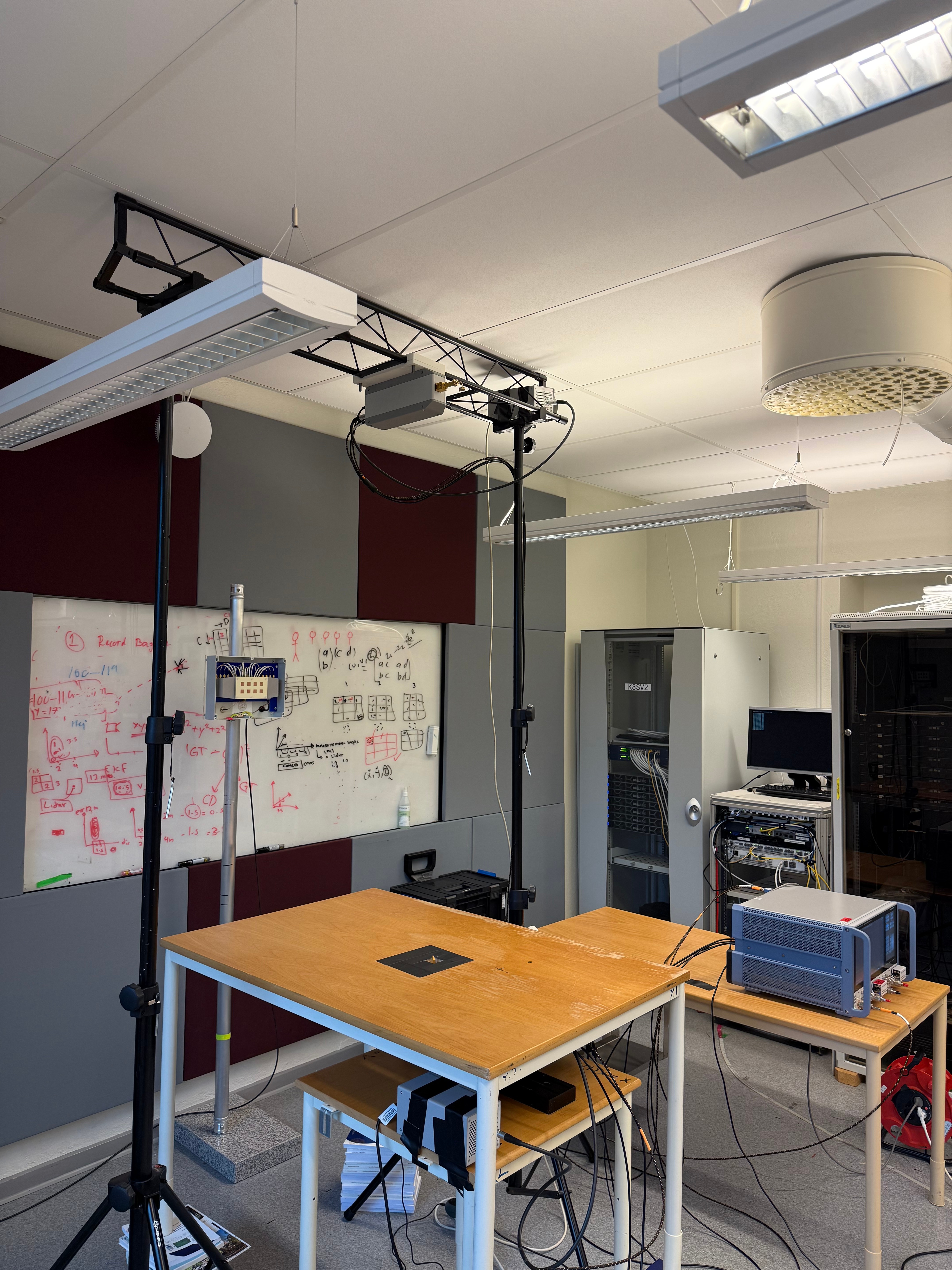}};
    \begin{scope}[x={(img.south east)}, y={(img.north west)}]
      \useasboundingbox (0,0) rectangle (1,1);

      \node[annot] (hole) at (0.75, 0.35) {30 X 30 mm\\opening};
      \draw[arrow] (hole.south) -- (0.6, 0.2);

      \node[annot] (desk) at (0.25, 0.1) {Shielding desk};
      \draw[arrow] (desk.east) -- (0.6, 0.15);

      \node[annot] (tx) at (0.76, 0.82) {Tx};
      \draw[arrow] (tx.west) -- (0.635, 0.77);

      \node[annot] (rx) at (0.45, 0.23) {Rx};
      \draw[arrow] (rx.east) -- (0.585, 0.2);

      \node[annot] (ceiling) at (0.25, 0.75) {Acoustic\\ceiling tiles};
      \draw[arrow] (ceiling.north) -- (0.5, 0.95);

    \end{scope}
  \end{tikzpicture}
  \label{fig:enviornment_5glab2}
  }
  \caption{Laboratory measurement environment.}
  \label{fig:setup_5GLab}
\end{figure}
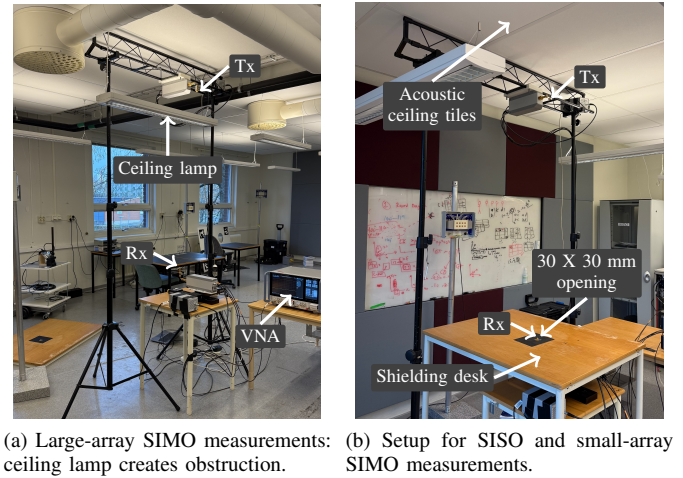

The second measurement environment was the 5G laboratory at EIT, Lund University, as shown in Fig.~\ref{fig:setup_5GLab}. In this environment, the large-array \gls{simo} measurements were performed at the location shown in Fig.~\ref{fig:enviornment_5glab1}, whereas the \gls{siso} and small-array \gls{simo} measurements were carried out at the location shown in Fig.~\ref{fig:enviornment_5glab2}.

The laboratory is characterized by exposed ventilation ducts, suspended cable trays, ceiling-mounted lighting fixtures, metallic mounting frames, measurement equipment racks, and suspended acoustic ceiling tiles likely made of glass wool. For the large-array \gls{simo} measurements, the \gls{tx} and \gls{rx} were intentionally arranged such that part of the synthesized array aperture was partially obstructed by a hanging ceiling lamp, as shown in Fig.~\ref{fig:enviornment_5glab1}.

\subsubsection{Ventilation room }
\begin{figure}[hbt!]
  \centering
  \subfloat[Large-array \gls{simo} measurements: metallic cable tray creates obstruction.]{
  \begin{tikzpicture}
    \node[anchor=south west, inner sep=0] (img) 
      {\includegraphics[width=.46\columnwidth]{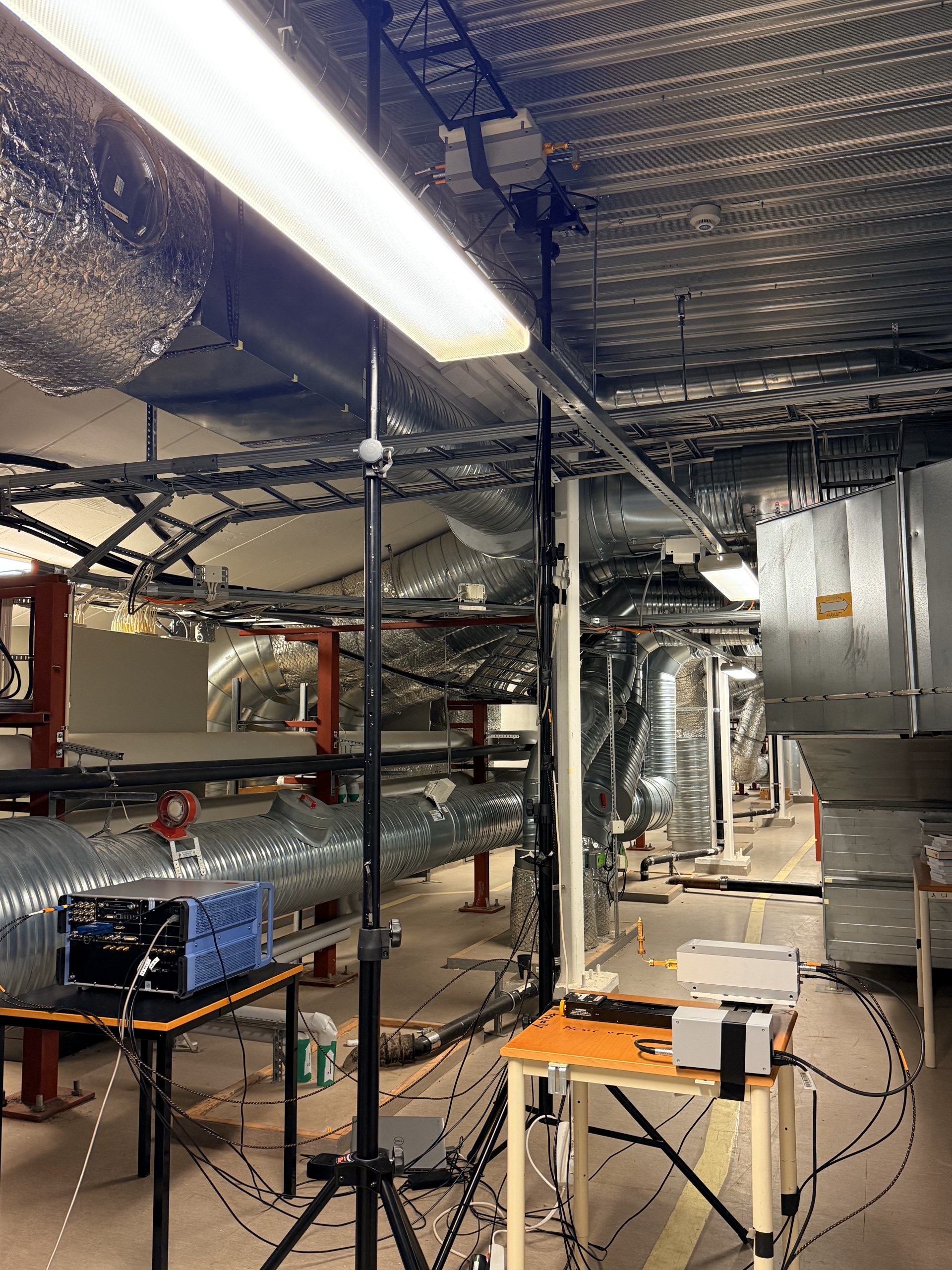}};
    \begin{scope}[x={(img.south east)}, y={(img.north west)}]
      \useasboundingbox (0,0) rectangle (1,1);

      \node[annot] (tray) at (0.3, 0.55) {Metallic\\cable tray};
      \draw[arrow] (tray.north) -- (0.6, 0.68);

      \node[annot] (tx) at (0.8, 0.9) {Tx};
      \draw[arrow] (tx.west) -- (0.61, 0.87);

      \node[annot] (rx) at (0.8, 0.32) {Rx};
      \draw[arrow] (rx.west) -- (0.68, 0.28);

      \node[annot] (vna) at (0.4, 0.32) {VNA};
      \draw[arrow] (vna.west) -- (0.25, 0.28);

      \node[annot] (ceiling) at (0.3, 0.75) {Corrugated\\metallic ceiling};
      \draw[arrow] (ceiling.east) -- (0.75, 0.8);

    \end{scope}
  \end{tikzpicture}
  \label{fig:enviornment_venroom1}
  }
  \hfill
  \subfloat[Setup for \gls{siso} and small-array \gls{simo} measurements.]
  {
  \begin{tikzpicture}
    \node[anchor=south west, inner sep=0] (img) 
      {\includegraphics[width=.46\columnwidth]{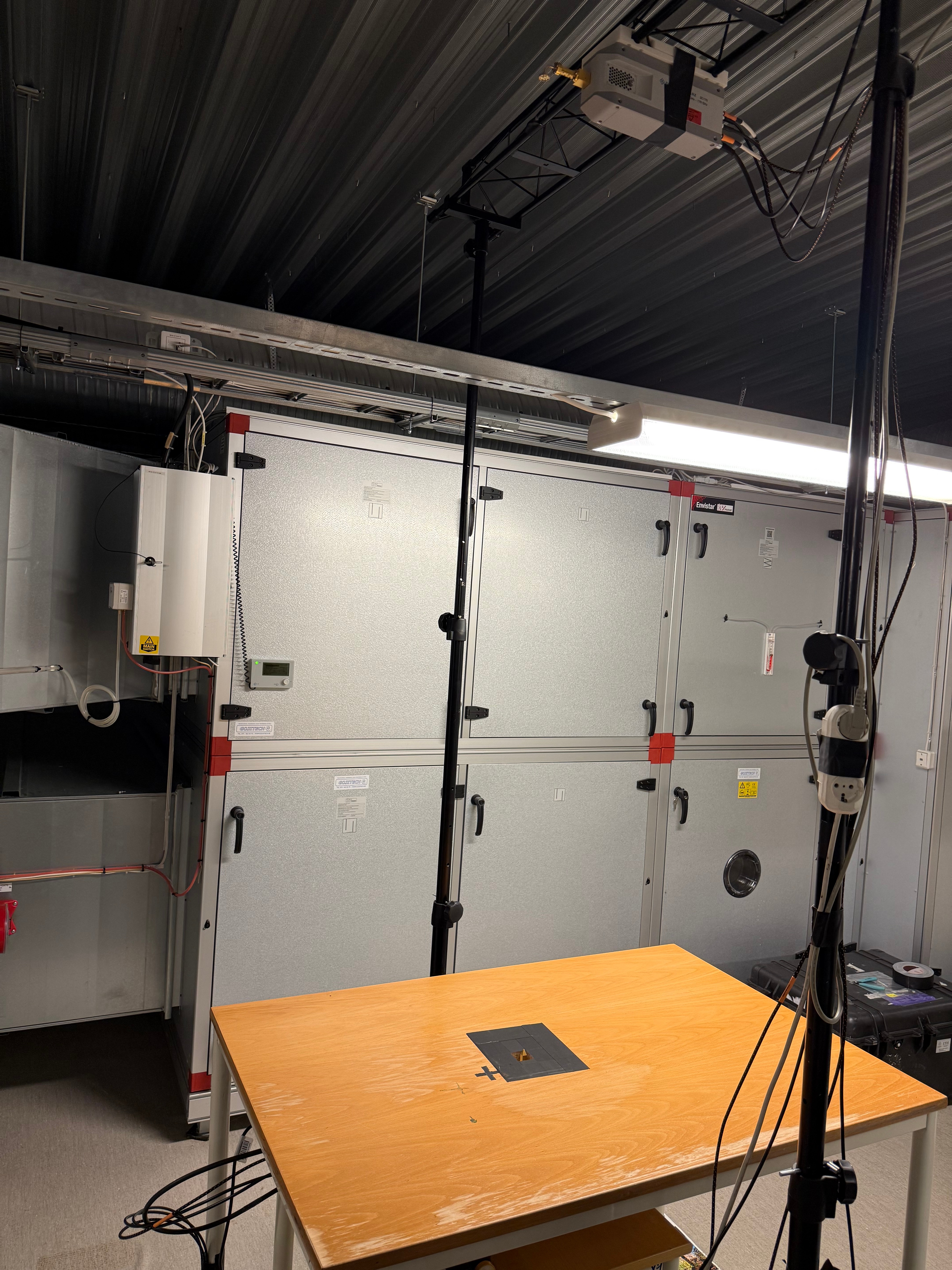}};
    \begin{scope}[x={(img.south east)}, y={(img.north west)}]
      \useasboundingbox (0,0) rectangle (1,1);

      \node[annot] (tx) at (0.4, 0.92) {Tx};
      \draw[arrow] (tx.east) -- (0.55, 0.93);

      \node[annot] (desk) at (0.25, 0.3) {Shielding desk};
      \draw[arrow] (desk.east) -- (0.6, 0.22);

      \node[annot] (rx) at (0.4, 0.14) {Rx};
      \draw[arrow] (rx.east) -- (0.54, 0.165);

      \node[annot] (ceiling) at (0.25, 0.67) {Corrugated\\metallic ceiling};
      \draw[arrow] (ceiling.north) -- (0.25, 0.85);

    \end{scope}
  \end{tikzpicture}
  \label{fig:enviornment_venroom2}
  }
  \caption{Ventilation room measurement environment.}
  \label{fig:enviornment_venroom}
\end{figure}
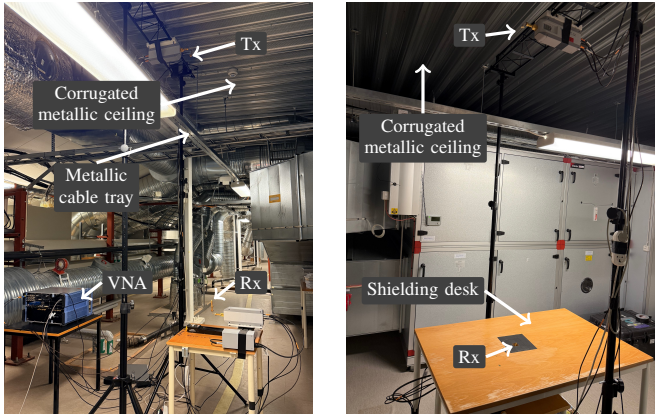

Finally, to emulate a more industrial-like propagation environment, measurements were conducted in a ventilation room located in the attic of the EIT building, Lund University, as shown in Fig.~\ref{fig:enviornment_venroom}. In this environment, the \gls{siso} and small-array \gls{simo} measurements were carried out at the location shown in Fig.~\ref{fig:enviornment_venroom2}, whereas the large-array \gls{simo} measurements were conducted at the location shown in Fig.~\ref{fig:enviornment_venroom1}. For the large-array case, the \gls{tx} and \gls{rx} were arranged such that part of the receiver array was obstructed by a metallic cable tray.

The ventilation room is a highly reflective and structurally complex environment, characterized by a corrugated metallic ceiling, low-hanging ceiling lights, metallic walls with handles and piping, and exposed ventilation ducts and equipment.

\subsection{Measurement Results}

\begin{figure}[h]
    \centering
    \includegraphics[width=1\linewidth]{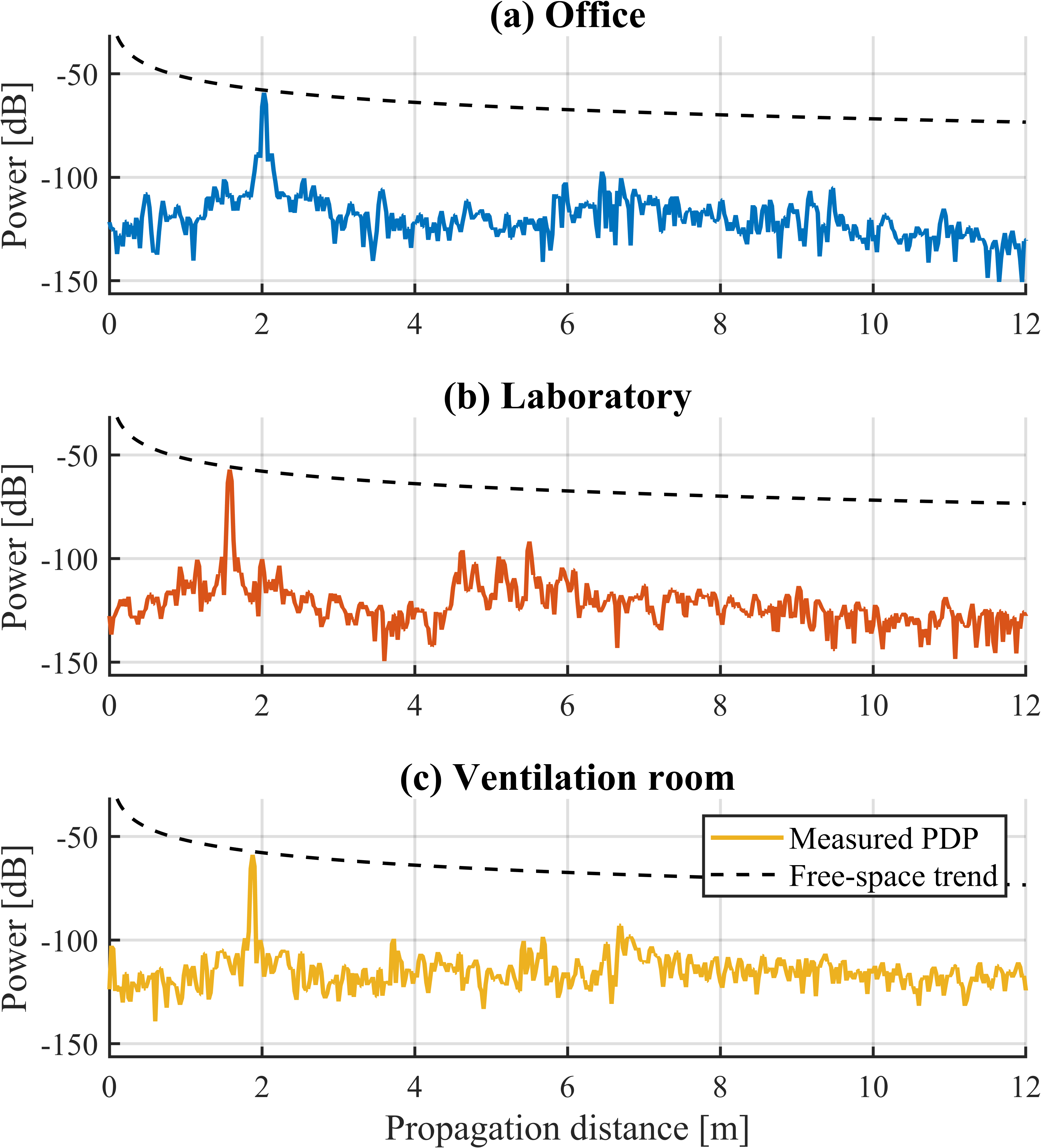}
    \caption{\glspl{pdp} obtained from \gls{siso} measurements.}
    \label{fig:SISO_PDPs}
\end{figure}

We first examine the \gls{siso} results shown in Fig.~\ref{fig:SISO_PDPs}, where the measured \glspl{pdp} for the three environments are plotted together with the theoretical free-space path-loss trend. 
In all environments, the strongest peak corresponds to the \gls{los} component. This peak appears at approximately 2~m propagation distance in the office, and at slightly shorter distances in the laboratory and ventilation room due to their lower ceiling heights. Weaker delayed peaks are also visible in all environments, showing that the vertical channel is not purely free-space even under \gls{los} propagation conditions.

In all environments, the \glspl{pdp} show second-order reflections in addition to the direct \gls{los} component. Here, second-order refers to paths that first reflect from the table surface surrounding the \gls{rx}, then interact with the ceiling, \gls{tx} structure, or nearby ceiling-mounted objects, before returning to the \gls{rx}. In the office, delayed peaks around 6--7~m are consistent with such table--ceiling--receiver paths, with additional interaction from the hanging lamps and metallic mesh ceiling visible in Fig.~\ref{fig:setup_combined}. In the laboratory, the main delayed components around 4.5--5.5~m agree with the shorter ceiling height and the geometry in Fig.~\ref{fig:setup_5GLab}, where reflections can involve the \gls{tx} structure, the acoustic ceiling tiles, and nearby ceiling-mounted infrastructure. In the ventilation room, stronger delayed components are visible around 5.5--7~m and over a wider propagation-distance range. These distances are consistent with the larger effective reflecting path created by the corrugated metallic ceiling and nearby metallic structures, including the cable tray, ducts, and walls shown in Fig.~\ref{fig:enviornment_venroom}. Similar effects, where metallic structures increase multipath richness and delay dispersion, have been reported in sub-THz industrial measurements~\cite{9838910}.

\begin{figure}[h]
    \centering
    \includegraphics[width=1\linewidth]{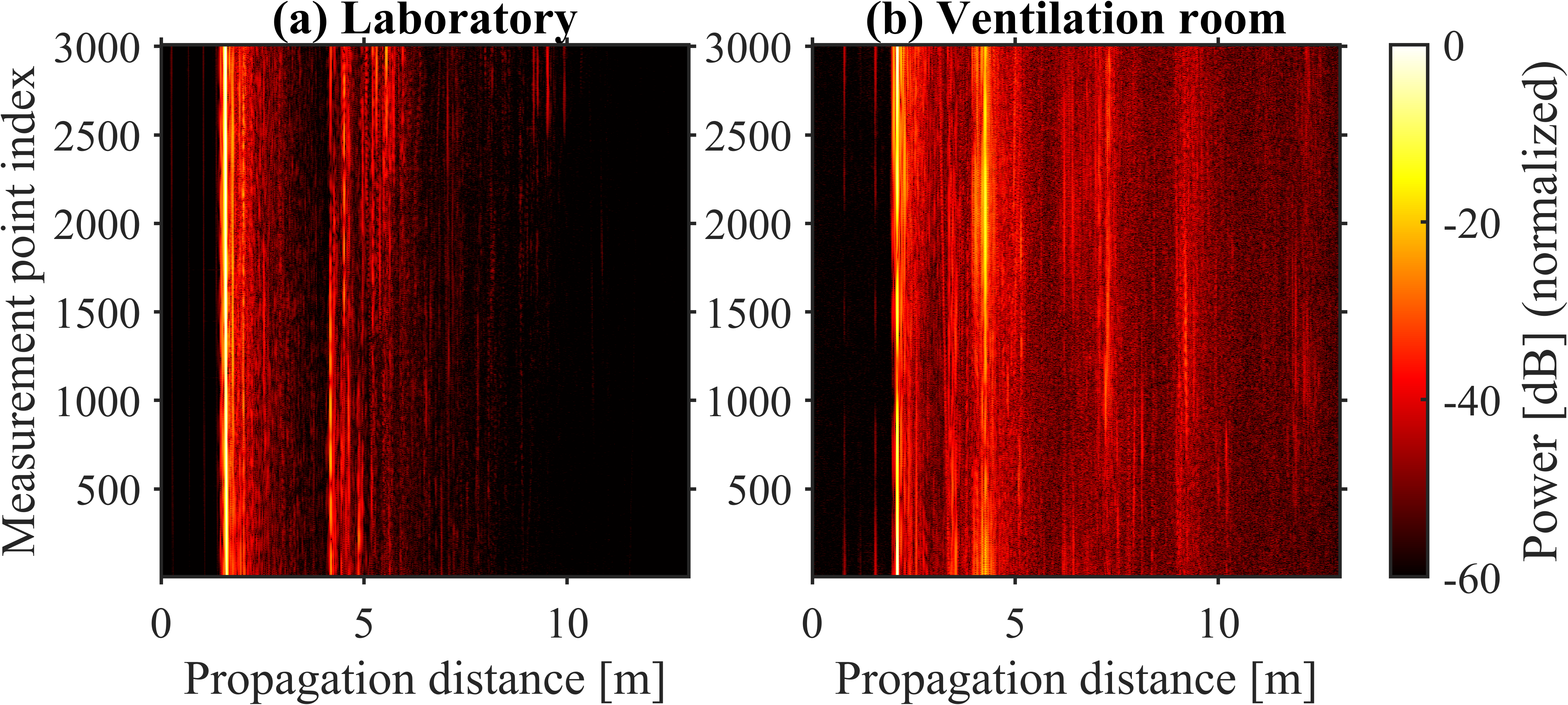}
    \caption{\glspl{pdp} from large-array \gls{simo} measurements.}
    \label{fig:large_Simo_PDPs}
\end{figure}

In the large-array \gls{simo} results shown in Fig.~\ref{fig:large_Simo_PDPs}, several propagation trajectories can be observed, including the \gls{los} component and multiple reflections. Many of these trajectories are visible only over limited portions of the array aperture, revealing clear spatial non-stationarity. For example, in the laboratory environment, some components around 10~m excess distance appear only over a small segment of the array, while other trajectories exhibit discontinuities caused by partial blockage along the array. The \gls{los} component also shows stronger power variation in the ventilation room than in the laboratory, indicating more severe obstruction, which is likely due to the metallic cable tray in the ventilation room compared with the less blocking fluorescent lamp structure in the laboratory. Consistent with the \gls{siso} results, the signal also persists over larger excess delays in the ventilation room, which is likely due to its more reflective environment with a larger amount of metal, including a corrugated metallic ceiling.

\begin{figure}[h]
  \centering
  \includegraphics[width=\columnwidth]{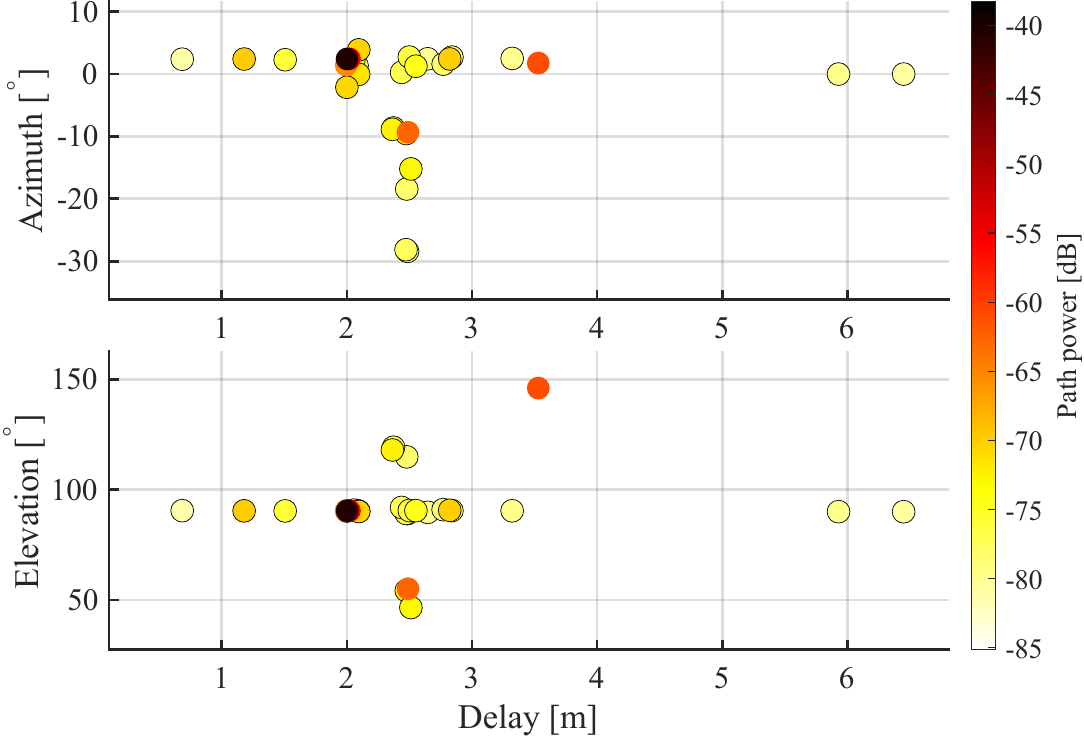}
  \caption{Estimated multipath components: Office.}
  \label{fig:sage_office}
\end{figure}

\begin{figure}[h]
  \centering
  \includegraphics[width=\columnwidth]{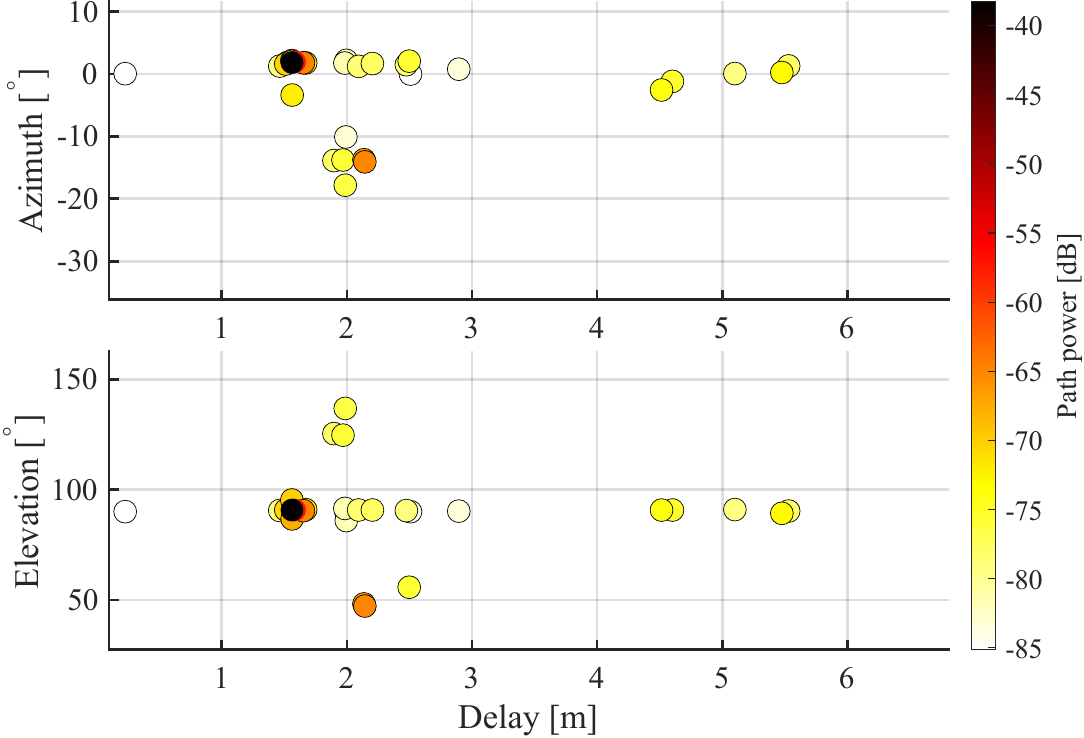}
  \caption{Estimated multipath components: Laboratory.}
  \label{fig:sage_lab}
\end{figure}

\begin{figure}[h]
  \centering
  \includegraphics[width=\columnwidth]{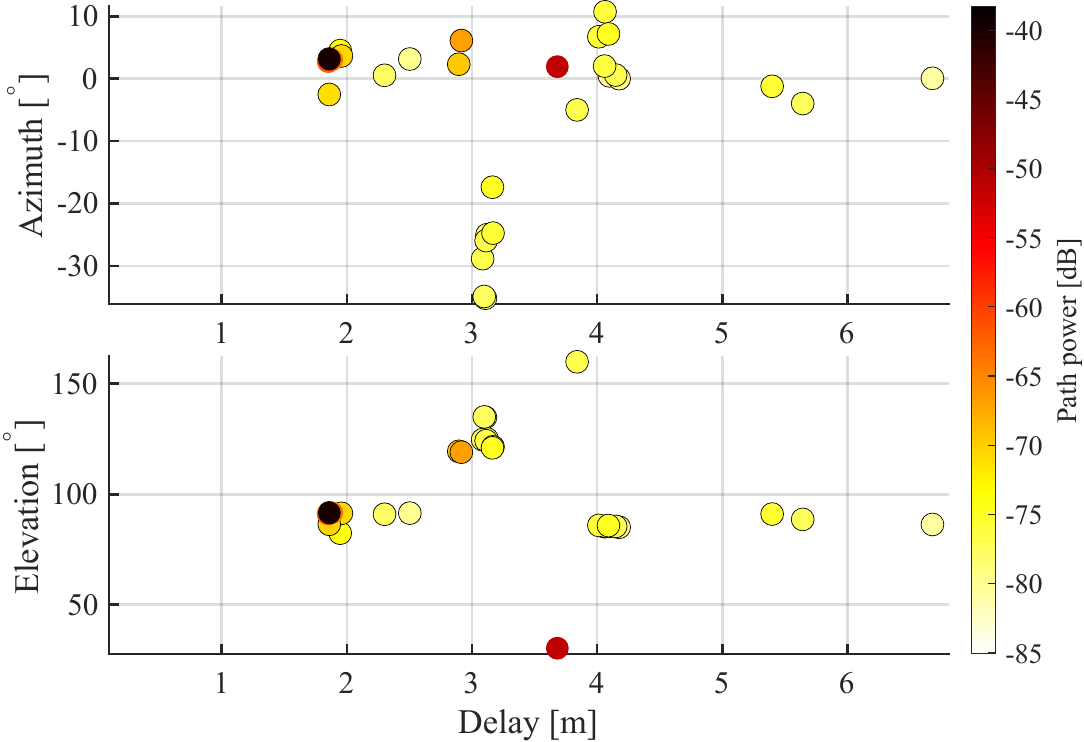}
  \caption{Estimated multipath components: Ventilation room.}
  \label{fig:sage_vent}
\end{figure}

For the small-array measurements, \gls{hrpe} was applied to the measured channel responses using the \gls{sage} algorithm \cite{753729}. This provides joint estimates of the delay, \gls{aoa}, and relative power of the dominant \glspl{mpc}. The resulting estimates for the office, laboratory, and ventilation-room environments are shown in Figs.~\ref{fig:sage_office}, \ref{fig:sage_lab}, and \ref{fig:sage_vent}, respectively, corresponding to the measurement positions in Figs.~\ref{fig:setup_combined}--\ref{fig:enviornment_venroom}.

The angular coordinates are defined with respect to the center of the \gls{rx} array. An elevation angle of $90^\circ$ corresponds to a wave arriving from directly above the array, i.e., along the vertical ceiling-to-ground direction, while elevations below and above $90^\circ$ represent arrivals from opposite sides of this vertical direction. The azimuth angle describes the horizontal orientation around the array, with $0^\circ$ corresponding to the nominal vertical plane containing the \gls{tx} and \gls{rx}. Hence, the \gls{los} component is expected to lie close to $0^\circ$ azimuth and $90^\circ$ elevation. Small deviations from this direction are observed in all three environments, mainly due to imperfect alignment between the \gls{tx} and the center of the \gls{rx} array.

In the office environment, Fig.~\ref{fig:sage_office} shows a rich set of resolvable \glspl{mpc}. The strongest component appears close to the expected vertical direction at a delay of approximately $2$~m and is identified as the \gls{los} path. Additional reflected paths are observed around $135^\circ$ and $50^\circ$ elevation, which are consistent with reflections from the hanging ceiling lamp and the metallic support pole, respectively, based on the geometry in Fig.~\ref{fig:setup_combined}. Two delayed components are also observed at approximately $5.9$--$6.4$~m. These are interpreted as second-order reflections involving the \gls{tx} frequency converter and the ceiling. Their separation of approximately $0.5$~m agrees with the additional round-trip distance between the \gls{tx} and the ceiling, since the \gls{tx} was located about $0.25$~m below the ceiling. The ceiling related second-order reflection is approximately $40.1$~dB weaker than the \gls{los} component, while the lamp reflection is about $32$~dB weaker.

In the laboratory environment, Fig.~\ref{fig:sage_lab}, the strongest component is again close to the expected \gls{los} direction, but appears at a shorter delay of approximately $1.56$~m due to the smaller \gls{tx}--\gls{rx} separation. The main second-order components occur between approximately $4.6$ and $5.5$~m. The components around $4.6$~m and $5.1$~m are attributed to reflections from the \gls{tx} frequency converter and the ceiling, respectively, with the latter being approximately $41.1$~dB weaker than the \gls{los} component. A further component around $5.5$~m arrives from a similar angular direction as the ceiling-reflected component, suggesting additional reflection or scattering from material or structures located inside or above the suspended acoustic ceiling tiles. This indicates that, for ceiling-mounted deployments, the ceiling material and internal ceiling structure may need to be considered when the ceiling allows partial penetration of the sub-THz wave, since layered or suspended ceilings can introduce additional reflected or scattered \glspl{mpc}.

Finally, in the ventilation room, Fig.~\ref{fig:sage_vent}, the \gls{los} component is observed at approximately $1.86$~m. Compared with the other environments, several stronger reflected \glspl{mpc} are visible, consistent with the large number of metallic structures in Fig.~\ref{fig:enviornment_venroom}, including the corrugated metallic ceiling, metallic walls, exposed ducts, and other installations. The delayed components include a reflection from the \gls{tx} structure around $5.5$~m and a ceiling-related second-order reflection around $6.7$~m, with the latter being approximately $40.7$~dB weaker than the \gls{los} component. The larger delay separation between these two delayed components, compared with the office and laboratory cases, is consistent with the larger effective distance between the \gls{tx} and the reflecting ceiling surface caused by the corrugated ceiling geometry.

The spread statistics in Tables~\ref{tab:smallarray_hybrid_spreads} and~\ref{tab:largearray_hybrid_spreads} summarize the \gls{simo} measurements in terms of mean delay, delay spread, and angular spread. Here, $\bar{\tau}$ denotes the mean delay, while $\sigma_{\tau}$ denotes the corresponding \gls{rms} delay spread. For both array configurations, these delay statistics are computed from spatially averaged, back-to-back-calibrated measured \glspl{pdp} obtained using a Hann-windowed IFFT and a 30~dB dynamic-range threshold relative to the \gls{pdp} peak. The averaging is performed over the $20\times20$ receiver positions for the small-array case and over the full $300\times10$ aperture for the large-array case. Since the \gls{pdp} does not provide angular information, the angular spreads are computed from the $K=30$ strongest \gls{sage}-estimated \glspl{mpc}. For the large-array case, $\mu_b(\sigma_{\phi,b})$ and $\mu_b(\sigma_{\theta,b})$ denote the mean of the blockwise azimuth and elevation spreads across blocks.

For the small-array measurements in Table~\ref{tab:smallarray_hybrid_spreads}, the office and laboratory environments show similar \gls{rms} delay spreads of $0.55$~ns and $0.59$~ns, respectively, whereas the ventilation room exhibits a much larger value of $1.74$~ns. This agrees with the \gls{siso} \glspl{pdp} and the \gls{sage} estimates, where the ventilation room shows stronger delayed multipath due to its corrugated metallic ceiling and surrounding metallic structures.

The angular spreads show a similar environment dependence. In azimuth, the spreads remain close to $1^\circ$ in all three environments, indicating that the dominant energy remains concentrated around the vertical plane containing the \gls{tx} and \gls{rx}. The elevation spread, however, increases from $2.65^\circ$ in the laboratory and $5.96^\circ$ in the office to $15.31^\circ$ in the ventilation room. This indicates that the additional multipath in the ventilation room is not only delayed, but also arrives over a wider range of vertical directions, consistent with reflections from the corrugated ceiling, walls, ducts, and other metallic structures.

\begin{table}[t]
\centering
\caption{\gls{rms} delay and angular spreads for the small-array \gls{simo} measurements.}
\label{tab:smallarray_hybrid_spreads}
\normalsize
\setlength{\tabcolsep}{6pt}
\renewcommand{\arraystretch}{1.15}
\begin{tabular}{lccc}
\toprule
Quantity & Office & Laboratory & Ventilation room \\
\midrule
$\bar{\tau}$ [ns]             & 6.75 & 5.24 & 6.74  \\
$\sigma_{\tau}$ [ns]          & 0.55 & 0.59 & 1.74  \\
$\sigma_{\phi}$ [$^\circ$]    & 1.23 & 0.98 & 1.13  \\
$\sigma_{\theta}$ [$^\circ$]  & 5.96 & 2.65 & 15.31 \\
\bottomrule
\end{tabular}
\end{table}

\begin{figure}[h]
  \centering
  \includegraphics[width=\columnwidth]{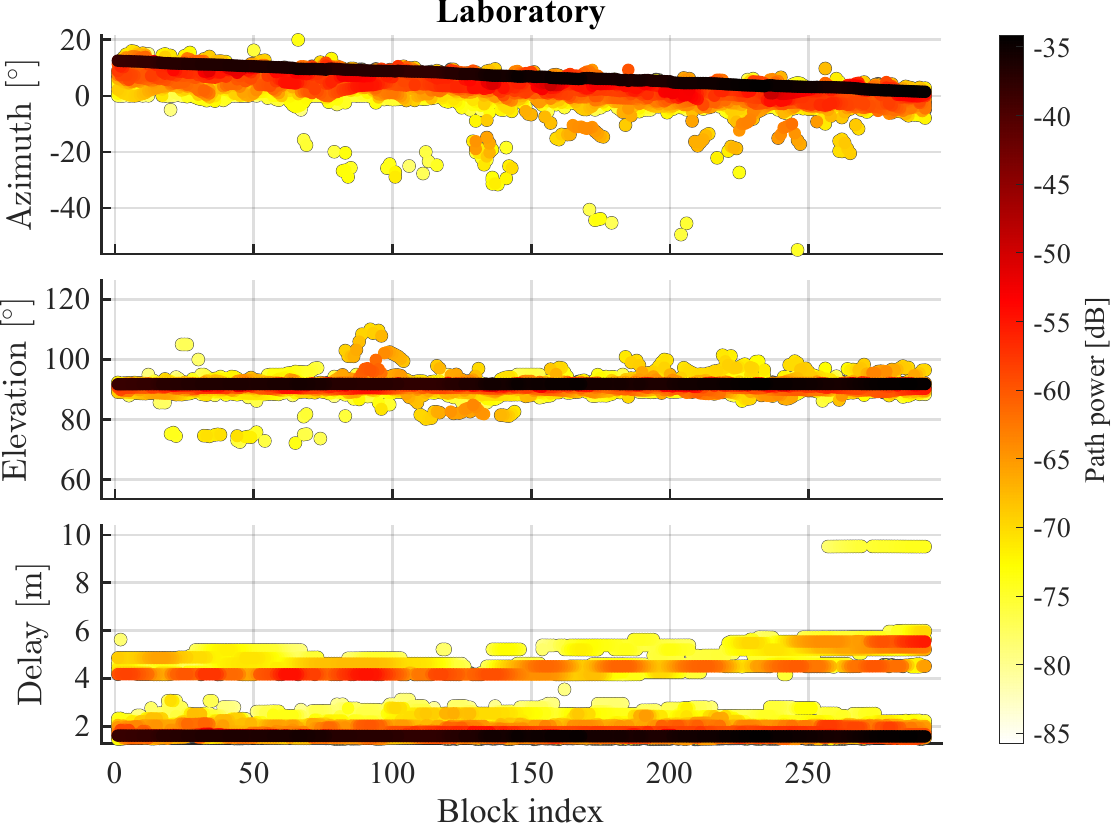}
  \caption{Concatenated MPCs for the laboratory environment estimated along the 30 cm long array, large-array \gls{simo} measurement.}
  \label{fig:large_sage_lab}
\end{figure}

\begin{figure}[h]
  \centering
  \includegraphics[width=\columnwidth]{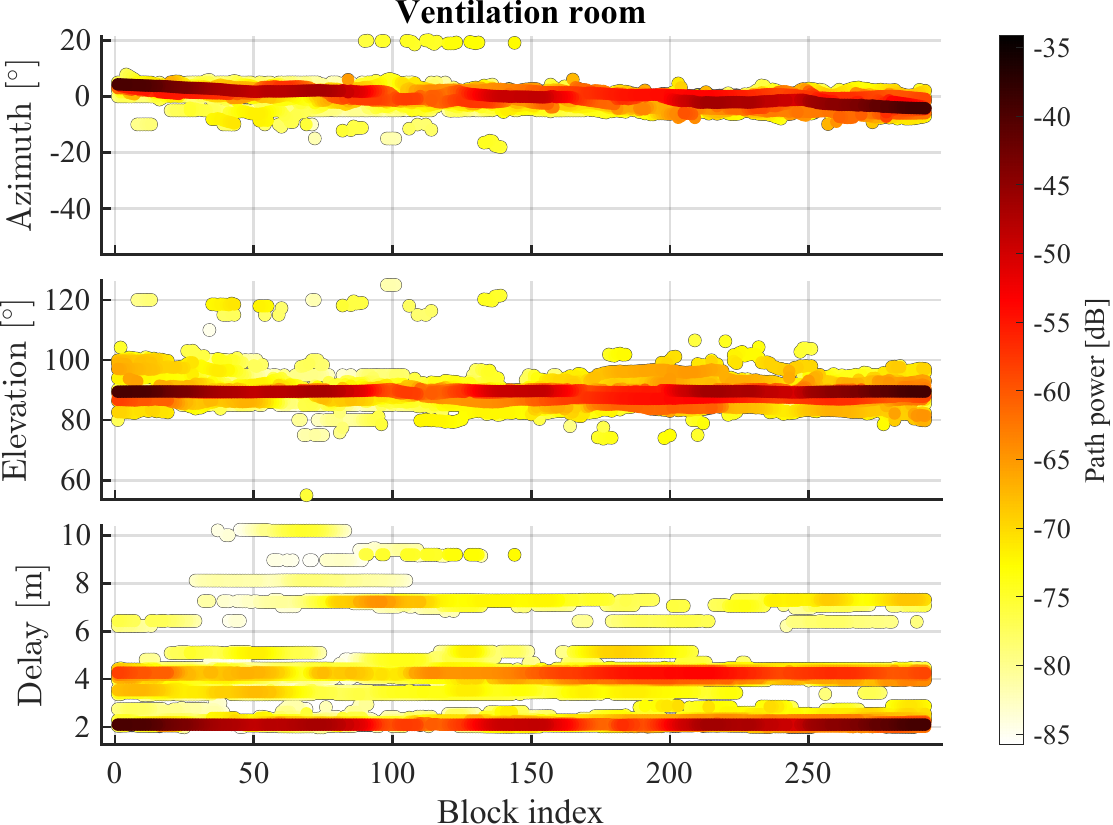}
  \caption{Concatenated MPCs for the ventilation room environment estimated along the 30 cm long array, large-array \gls{simo} measurement.}
  \label{fig:large_sage_vent}
\end{figure}

It should be noted that the large-array measurements were conducted at different positions from the corresponding \gls{siso} and small-array \gls{simo} measurements in the laboratory and ventilation-room environments, as shown in Figs.~\ref{fig:setup_5GLab} and~\ref{fig:enviornment_venroom}. The large-array \gls{simo} measurements provide insight into spatial non-stationarity along the extended aperture. Figs.~\ref{fig:large_sage_lab} and~\ref{fig:large_sage_vent} show the concatenated \gls{sage}-estimated \glspl{mpc} along the 30~cm virtual array in the laboratory and ventilation room, respectively. Several paths are visible only over limited block ranges, and some trajectories appear or disappear along the aperture, showing that a single stationary set of \glspl{mpc} is not sufficient to describe the full large-array channel.

In the laboratory, Fig.~\ref{fig:large_sage_lab} shows a dominant component over most of the aperture, with weaker delayed paths appearing only over shorter block intervals. In the ventilation room, Fig.~\ref{fig:large_sage_vent} shows stronger path-power variations and more persistent delayed components, consistent with the large-array \gls{pdp} in Fig.~\ref{fig:large_Simo_PDPs}. This is attributed to the metallic cable tray and corrugated metallic ceiling, which introduce both blockage and additional reflected paths.

Table~\ref{tab:largearray_hybrid_spreads} summarizes the large-array spread statistics. The ventilation room has a much larger \gls{rms} delay spread than the laboratory, increasing from $0.44$~ns to $2.57$~ns. The mean delay also increases from $5.33$~ns to $7.99$~ns, indicating overall more delayed energy.

The angular spreads are much smaller than the delay-spread differences. This is partly because the large-array angular spreads are computed as mean blockwise \gls{sage}-based spreads, which describe local angular dispersion within each block rather than the angular variation over the full aperture. In addition, the large-array measurements used a more directive \gls{tx} horn antenna, while the small-array measurements used a broader open-waveguide \gls{tx}. Therefore, the large-array results mainly show stronger delay dispersion and aperture-dependent visibility changes, while the dominant angular energy remains concentrated near the vertical ceiling-to-ground direction.

\begin{table}[t]
\centering
\caption{\gls{rms} delay and angular spreads for the large-array \gls{simo} measurements.}
\label{tab:largearray_hybrid_spreads}
\normalsize
\setlength{\tabcolsep}{7pt}
\renewcommand{\arraystretch}{1.15}
\begin{tabular}{lcc}
\toprule
Quantity & Laboratory & Ventilation room \\
\midrule
$\bar{\tau}$ [ns]                       & 5.33 & 7.99 \\
$\sigma_{\tau}$ [ns]                    & 0.44 & 2.57 \\
$\mu_b(\sigma_{\phi,b})$ [$^\circ$]     & 0.89 & 1.12 \\
$\mu_b(\sigma_{\theta,b})$ [$^\circ$]   & 0.24 & 1.15 \\
\bottomrule
\end{tabular}
\end{table}

To place these values in context, Table~\ref{tab:comparison_subthz} compares the measured \gls{rms} delay spreads with selected indoor sub-THz campaigns in more conventional horizontal-link geometries. Compared with reported office, factory, and industrial-like open-office measurements at similar frequencies, the vertical links in this work show smaller \gls{rms} delay spreads, with ranges of $0.55$--$1.74$~ns for the small-array measurements and $0.44$--$2.57$~ns for the large-array measurements. This is mainly due to the short ceiling-to-ground geometry where the channel is strongly dominated by the \gls{los} path. Such behavior is favorable for ceiling-mounted \glspl{ru}, as envisioned in the 6GTandem concept, since it can provide a strong link budget, reduced delay dispersion, and potentially allow broader beams than long horizontal links without collecting excessive multipath.

However, as seen from Figs.~\ref{fig:sage_office}, \ref{fig:sage_lab}, and \ref{fig:sage_vent}, the vertical channel is not purely free-space. The measured \glspl{pdp} and \gls{sage}-estimated \glspl{mpc} show repeatable second-order reflections involving the table, ceiling, and \gls{tx} structure. In the office environment, reflections from the hanging ceiling lamps are clearly observed, while in the laboratory, the suspended acoustic ceiling appears to introduce additional reflected or scattered components from structures above or within the ceiling. The ventilation room further shows that corrugated metallic ceilings, cable trays, and other suspended metallic objects can significantly increase delay dispersion and aperture-dependent variations. Thus, vertical sub-THz links offer a more confined and \gls{los}-dominated propagation geometry, but ceiling materials, lamps, and other ceiling-mounted obstructions must still be considered in channel modeling, beam design, and deployment planning.

\begin{table}[t]
\centering
\caption{Comparison with selected indoor sub-THz measurements.}
\label{tab:comparison_subthz}
\normalsize
\setlength{\tabcolsep}{4pt}
\renewcommand{\arraystretch}{1.12}
\begin{tabular}{lcc}
\toprule
Scenario & Frequency & $\sigma_{\tau}$ [ns] \\
\midrule
This work, small-array vertical links & 140 GHz & 0.55--1.74 \\
This work, large-array vertical links & 140 GHz & 0.44--2.57 \\
Office, directional LOS~\cite{9450830} & 142 GHz & 2.71 \\
Office, omnidirectional LOS~\cite{9450830} & 142 GHz & 3.00 \\
Office, omnidirectional NLOS~\cite{9450830} & 142 GHz & 9.20 \\
Factory, directional~\cite{9838910} & 142 GHz & 3.0 \\
Factory, omnidirectional~\cite{9838910} & 142 GHz & 16.0 \\
Open office, LOS~\cite{11000040} & 160 GHz & 16.5 \\
Open office, NLOS~\cite{11000040} & 160 GHz & 28.4 \\
\bottomrule
\end{tabular}
\end{table}

\section{Conclusions} \label{sec:conclusions}

This paper presented a measurement-based characterization of indoor vertical ceiling-to-ground sub-THz channels in the 136--144~GHz band. A \gls{vna}-based channel sounder with a planar \gls{vaa} receiver was implemented and successfully used to capture \gls{siso}, small-array \gls{simo}, and large-array \gls{simo} channel responses in office, laboratory, and ventilation-room environments. The measurements demonstrate that the implemented sounder provides sufficient phase stability and spatial resolution to resolve dominant multipath components and aperture-dependent channel variations in vertical sub-THz links.

The results show that the investigated vertical links are generally dominated by a strong \gls{los} component close to the ceiling-to-ground direction. Compared with reported horizontal indoor sub-THz measurements, the measured \gls{rms} delay spreads are relatively small. This supports the potential of ceiling-mounted \glspl{ru}, as envisioned in the 6GTandem concept, where short vertical links can provide \gls{los} propagation and limited delay dispersion.

However, the vertical channel is not purely free-space. Repeatable second-order reflections involving the table, ceiling, and \gls{tx} structure were observed. In the laboratory, additional reflected or scattered components indicate that structures above or within the suspended acoustic ceiling can contribute to the channel when the ceiling allows partial penetration of the sub-THz wave. Ceiling-mounted objects and metallic structures, such as lamps, cable trays, and corrugated metallic ceilings, further introduced additional reflected components and aperture-dependent variations.

The large-array measurements demonstrated spatial non-stationarity along the 30~cm aperture, with several \glspl{mpc} visible only over limited parts of the array. These findings indicate that vertical sub-THz links can benefit from short link distances and confined propagation, but that ceiling materials, internal ceiling structures, and suspended obstructions must be considered in channel modeling, beam design, equalizer design, and deployment planning.

\section{ACKNOWLEDGMENT} \label{sec:ACKNOWLEDGMENT}
This work was supported by the Swedish Research Council (Grant No. 2022-04691), the Horizon Europe Programme (Marie Skłodowska-Curie Grant No. 101059091), the Horizon 2020 Programme (Grant No. 861222), the Royal Physiographic Society of Lund, 6GTandem, ELLIIT, and Ericsson.

\bibliographystyle{IEEEtran}
\bibliography{ref}

\end{document}